\documentclass[twocolumn,epjc3]{svjour3}
\journalname{}
\usepackage[utf8]{inputenc}
\usepackage[T1]{fontenc}

\usepackage[numbers,sort&compress]{natbib}

\usepackage{graphicx}
\usepackage{scalefnt}

\usepackage{cuted}
\usepackage{flushend}

\usepackage{amsmath}
\usepackage{amssymb}
\usepackage{amsfonts}
\usepackage{mathtools}
\usepackage{xpatch}

\usepackage{txfonts}
\usepackage{microtype}

\usepackage[dvipsnames]{xcolor}

\usepackage{enumitem}
\newlist{renumerate}{enumerate}{1}
\setlist[renumerate]{label=(\roman*),leftmargin=*}
\newlist{aenumerate}{enumerate}{1}
\setlist[aenumerate]{label=(\arabic*),leftmargin=*}

\usepackage[hyphens]{url}
\usepackage{hyperref}
\usepackage[hyphenbreaks]{breakurl}

\usepackage[frozencache=true,cachedir=minted-cache]{minted} 
\usepackage{listings}

\newcommand{\bra}{\langle}
\newcommand{\ket}{\rangle}
\newcommand{\nuhamil}{\texttt{NuHamil} }
\newcommand{\neworrenewcommand}[1]{\providecommand{#1}{}\renewcommand{#1}}
\renewcommand{\vec}[1]{\boldsymbol{#1}}

\usepackage{xurl}

\newcommand{\cg}[6]{\mathcal{C}^{#1#2#3}_{#4#5#6}}

\newcommand{\sixj}[6]{\left\{
    \begin{array}{ccc}
      #1 & #2 & #3 \\
      #4 & #5 & #6
    \end{array}
\right\}}

\newcommand{\ninej}[9]{\left\{
    \begin{array}{ccc}
      #1 & #2 & #3 \\
      #4 & #5 & #6 \\
      #7 & #8 & #9
    \end{array}
\right\}}

\newcommand{\twlvj}[9]{
  \neworrenewcommand{\ffoo}[3]{\left\{
    \begin{array}{cccccccc}
      #1 &    & #2  &    & #3  &    & #4  &    \\
         & #5 &     & #6 &     & #7 &     & #8 \\
      #9 &    & ##1 &    & ##2 &    & ##3 &
    \end{array}
\right\}
    }
    \ffoo
}

\begin{document}\sloppy

\allowdisplaybreaks

\title{\nuhamil: A numerical code to generate nuclear two- and three-body matrix elements from chiral effective field theory}

\author{Takayuki Miyagi\thanksref{aff1,aff2,aff3,em:at}}
\thankstext{em:at}{\email{miyagi@theorie.ikp.physik.tu-darmstadt.de}}
\institute{%
\label{aff1}%
Technische Universit\"at Darmstadt, Department of Physics, 64289 Darmstadt, Germany
\and
\label{aff2}
ExtreMe Matter Institute EMMI, GSI Helmholtzzentrum f\"ur Schwerionenforschung GmbH, 64291 Darmstadt, Germany
\and
\label{aff3}
Max-Planck-Institut f\"ur Kernphysik, Saupfercheckweg 1, 69117 Heidelberg, Germany
}


\maketitle
\begin{abstract}
The applicability of nuclear {\it ab initio} calculations has rapidly extended over the past decades. However, starting research projects is still challenging due to the required numerical expertise in the generation of underlying nuclear interaction matrix elements and many-body calculations. To ease the first issue, in this paper we introduce the numerical code \nuhamil to generate the nucleon-nucleon (NN) and three-nucleon (3N) matrix elements expressed in a spherical harmonic-oscillator basis, inputs of many-body calculations. The ground-state energies for the selected doubly closed shell nuclei are calculated with the no-core shell-model (NCSM) and in-medium similarity renormalization group (IMSRG). The code is written in modern Fortran, and OpenMP+MPI hybrid parallelization is available for the 3N matrix-element calculations.
\end{abstract}
%

\section*{PROGRAM SUMMARY}
\begin{description}[font=\normalfont\itshape]
  \item[Program title:] \texttt{NuHamil}
  \item[Licensing provisions:] GPLv3
  \item[Programming language:] Modern Fortran
  \item[Repository and DOI:]
  \url{https://github.com/Takayuki-Miyagi/NuHamil-public}\\
  DOI:
  \url{https://doi.org/10.5281/zenodo.7529481}
  \item[Description of problem:] Nucleon-nucleon (NN) and three-nucleon (3N) matrix elements are essential inputs in nuclear {\it ab initio} calculations.
  However, developing a numerical code to generate the matrix elements is a demanding task.
Preparing the input matrix elements is one of the main barriers to begin studies.
  \item[Method of solution:] The \nuhamil code has the capability to generate both NN and 3N matrix elements expressed in a single-particle harmonic-oscillator (HO) basis, which can be used as inputs for most of the {\it ab initio} calculation methods. The jobs can be managed by a simple Python script.
  \item[Additional comments:]
 For other open-source software, one can use the computational environment for nuclear structure (CENS)~\cite{CENS} and recently published NuclearToolkit code~\cite{Yoshida2022}.
\end{description}

\section{Introduction}\label{sec1}
The dynamics of the atomic nucleus are governed by the strong interaction, whose fundamental theory is described by quantum chromodynamics (QCD).
Since the quarks are tightly confined in a nucleon, it is well established that nuclear Hamiltonians associated with the interactions between nucleons are a good starting point to understand nuclear structure and reactions.
The study of nuclear interaction models has a long story starting from the pion-exchange theory~\cite{Machleidt2017}.
The quantitative understanding of nuclear interactions is an open problem.
In the past decade, interactions based on chiral effective-field theory (EFT)~\cite{Epelbaum2009,Machleidt2011} have become the standard starting point of {\it ab initio} many-body calculations.
Chiral EFT-based interactions have several advantages over other interactions such as AV18~\cite{Wiringa1995} and CD-Bonn potentials~\cite{Machleidt2001}.
For example, a systematic expansion is possible by ordering the diagrams according to the power counting, suggesting the possibility of an uncertainty quantification due to the truncation in the expansion~\cite{Epelbaum2015,Furnstahl2015a,Melendez2017,Melendez2019}.
Further, many-nucleon interactions naturally appear at higher order, explaining the hierarchy of many-body terms.
In nuclear physics, it is well known that 3N interactions play an important role (see for example Ref.~\cite{Hebeler2021} as a recent review).
With the progress in nuclear interactions and methodological developments in many-body problems, nuclear {\it ab initio} studies are well motivated.
Nowadays, the applicability extends over the nuclear chart~\cite{Hergert2020} and recently reached to the heaviest known doubly-magic system $^{208}$Pb~\cite{Hu2021}, and further applications are expected.

To perform {\it ab initio} calculations, the matrix elements of nuclear Hamiltonians (and relevant operators) are essential.
However, developing a numerically efficient code for the matrix-element generation requires expert knowledge and can be a barrier for those entering the field.
The goal of the \nuhamil code is to provide a simple way to generate NN and 3N matrix elements expressed in a spherical HO basis, applicable for basis-expansion methods, such as the no-core shell model (NCSM)~\cite{Barrett2013}, coupled-cluster method~\cite{Hagen2014}, self-consistent Green's function method~\cite{Soma2020a}, in-medium similarity renormalization group~\cite{Hergert2016} approach, and man-body perturbation theory~\cite{Tichai2020}.
The code currently supports the input formats of the open-source \texttt{BIGSTICK}~\cite{Johnson2018} and \texttt{imsrg++}~\cite{imsrg++} codes for the NCSM and IMSRG calculations, respectively.

This paper is organized as follows.
In Sec.~\ref{sec:Ham}, we clarify the input NN and 3N matrix elements for the many-body calculations and briefly show how to compute them with the single-particle product state in the HO basis.
Also, we discuss the free-space similarity renormalization group (SRG) prescription to soften the nuclear interactions in Sec.~\ref{sec:SRG}.
We show some benchmark many-body calculation results with NCSM and IMSRG calculations in Sec.~\ref{sec:Results}.
The usage of the code and the conclusion are given in Sec.~\ref{sec:Usage} and  Sec.~\ref{sec:Conclusion}, respectively.

\section{Matrix elements of Hamiltonian \label{sec:Ham}}
Here, we review how the NN and 3N matrix elements enter in the many-body problem.
Our numerical goal is to solve the non-relativistic many-body Schr\"odinger equation $H \vert \Psi_{n} \ket =  E_{n} \vert \Psi_{n} \ket$, with the intrinsic Hamiltonian with up to 3N terms
\begin{equation}
\label{eq:Hcl}
\begin{aligned}
H & = \frac{A-1}{A}  \sum_{i}  \frac{\vec{p}^{2}_{i}}{2m} + \sum_{i<j} (V^{{\rm NN}}_{ij} - T^{\rm NN}_{ij})
+ \sum_{i<j<k} V^{3N}_{ijk}.
\end{aligned}
\end{equation}
Here, $\vec{p}_{i}$ is the momentum vector of $i$th nucleon, and $m$ is the nucleon mass.
The factor $(A-1)/A$ in the first term and $T^{\rm NN}_{ij}$ is from the subtraction of the center-of-mass (cm) kinetic term. $A$ is the nucleon number of the system.
The terms $V^{\rm NN}_{ij}$ and $V^{\rm 3N}_{ijk}$ are the NN and 3N interactions, respectively.

\subsection{Second-quantized representation}
To proceed with many-body calculations with basis-expansion methods, we begin with the expression by the second quantization.
To this end, we first define creation and annihilation operators for a nucleon in an HO orbit $\tilde{p}$: $c^{\dag}_{\tilde{p}}$ and $c_{\tilde{p}}$.
The subscript $\tilde{p}$ is a collective index specifying the HO orbit and defined as $\tilde{p} = \{n_{p}, l_{p}, j_{p}, m_{p}, t_{z,p}\}$.
Here, $n_{p}$, $l_{p}$, $j_{p}$, $m_{p}$, and $t_{z,p}$ are the nodal quantum number, orbital angular momentum, total angular momentum, $z$-component of $j_{p}$, and the $z$-component of the isospin distinguishing protons and neutrons, respectively.
Note that proton (neutron) states are labeled as $t_{z}=-1/2$ $(1/2)$.
In nuclear physics, the use of an HO basis is particularly useful since the coordinate transformation coefficient is well known~\cite{Moshinsky1959,Trlifaj1972,Kamuntavicius2001}, as will be shown in later.
The creation and annihilation operators satisfy the anticommutation relations
\begin{equation}
\{c_{\tilde{p}}, c_{\tilde{q}}\} = 0, \,
\{c^{\dag}_{\tilde{p}}, c^{\dag}_{\tilde{q}} \} = 0, \,
\{c_{\tilde{p}}, c^{\dag}_{\tilde{q}} \} = \delta_{\tilde{p}\tilde{q}}.
\end{equation}
The object $\delta_{\tilde{p}\tilde{q}}$ is defined by products of Kronecker's delta and is written as
\begin{equation}
\delta_{\tilde{p}\tilde{q}} = \delta_{n_{p}n_{q}} \delta_{l_{p}l_{q}} \delta_{j_{p}j_{q}} \delta_{m_{p}m_{q}} \delta_{t_{z,p}t_{z,q}}.
\end{equation}
Applying the creation operators to the nucleon vacuum state $|0\ket$, one can define antisymmetrized states.
For example, one-, two-, and three-nucleon states can be written as
\begin{align}
|\tilde{p} \ket &= c^{\dag}_{\tilde{p}} | 0 \ket, \\
|\tilde{p}\tilde{q} \ket &= c^{\dag}_{\tilde{p}} c^{\dag}_{\tilde{q}} | 0 \ket, \\
|\tilde{p}\tilde{q}\tilde{r} \ket &= c^{\dag}_{\tilde{p}} c^{\dag}_{\tilde{q}}c^{\dag}_{\tilde{r}} | 0 \ket.
\end{align}
Using the creation and annihilation operators, an arbitrary $n$-body operator $O^{[n]}$ can be expressed as
\begin{equation}
O^{[n]} = \left(\frac{1}{n!}\right)^{2} \sum_{\tilde{p}'_{1} \cdots \tilde{p}'_{n}} \sum_{\tilde{p}_{1}\cdots \tilde{p}_{n}} O_{\tilde{p}'_{1}\cdots \tilde{p}'_{n} \tilde{p}_{1}\cdots \tilde{p}_{n}}
c^{\dag}_{\tilde{p}'_{1}} \cdots c^{\dag}_{\tilde{p}'_{n}} c_{\tilde{p}_{n}} \cdots c_{\tilde{p}_{1}}
\end{equation}
The object $O_{\tilde{p}'_{1}\cdots \tilde{p}'_{n} \tilde{p}_{1}\cdots \tilde{p}_{n}}$ is a shorthand notation for the operator matrix element $\bra \tilde{p}'_{1}\cdots \tilde{p}'_{n} | O^{[n]} | \tilde{p}_{1}\cdots \tilde{p}_{n} \ket$.
In the same way, the Hamiltonian in Eq.~\eqref{eq:Hcl} can be quantized as
\begin{equation}
\begin{aligned}
\label{eq:2nd_quantization}
H &= \sum_{\tilde{p}'\tilde{p}} T_{\tilde{p}'\tilde{p}}c^{\dag}_{\tilde{p}'}c_{\tilde{p}}
\\ & + \left(\frac{1}{2!}\right)^{2}\sum_{\tilde{p}'\tilde{q}'\tilde{p}\tilde{q}} (V^{\rm NN}_{\tilde{p}'\tilde{q}'\tilde{p}\tilde{q}}  - T^{\rm NN}_{\tilde{p}'\tilde{q}'\tilde{p}\tilde{q}}) c^{\dag}_{\tilde{p}'}c^{\dag}_{\tilde{q}'}c_{\tilde{q}}c_{\tilde{p}}
\\ & +
\left(\frac{1}{3!}\right)^{2}\sum_{\tilde{p}'\tilde{q}'\tilde{r}'\tilde{p}\tilde{q}\tilde{r}} V^{\rm 3N}_{\tilde{p}'\tilde{q}'\tilde{r}'\tilde{p}\tilde{q}\tilde{r}} c^{\dag}_{\tilde{p}'}c^{\dag}_{\tilde{q}'}c^{\dag}_{\tilde{r}'}c_{\tilde{r}}c_{\tilde{q}}c_{\tilde{p}},
\end{aligned}
\end{equation}
using the matrix elements of one-body kinetic term $T_{\tilde{p}'\tilde{p}}$, two-body kinetic term $T^{\rm NN}_{\tilde{p}'\tilde{q}'\tilde{p}\tilde{q}}$, NN interaction $V^{\rm NN}_{\tilde{p}'\tilde{q}'\tilde{p}\tilde{q}}$, and 3N interaction $V^{\rm 3N}_{\tilde{p}'\tilde{q}'\tilde{r}'\tilde{p}\tilde{q}\tilde{r}}$.

\subsection{$J$-coupled scheme}
The number of matrix elements defined in Eq.~\eqref{eq:2nd_quantization} is greatly reduced by exploiting the rotational symmetry of the Hamiltonian.
To introduce a smaller set of matrix elements, we define the $J$-coupled two- and three-body states using the Clebsch-Gordan coefficient $\cg{j_{1}}{j_{2}}{J}{m_{1}}{m_{2}}{m_{1}+m_{2}}$:
\begin{align}
|pq: JM \ket &= \sqrt{\frac{1}{1+\delta_{pq}}}\sum_{m_{p}m_{q}} \cg{j_{p}}{j_{q}}{J}{m_{p}}{m_{q}}{M}  |\tilde{p}\tilde{q} \ket, \\
|pqr: J_{pq}JM\ket &= \sum_{m_{p}m_{q}m_{r}} \cg{j_{p}}{j_{q}}{J_{pq}}{m_{p}}{m_{q}}{m_{p}+m_{q}}
\cg{J_{pq}}{j_{r}}{J}{m_{p}+m_{q}}{m_{r}}{M} | \tilde{p}\tilde{q}\tilde{r} \ket.
\end{align}
Here, $p$, $q$, and $r$ are the quantum number set without $m$, i.e. $p = \{n_{p}, l_{p}, j_{p}, t_{z,p}\}$, and an additional Kronecker's delta product is introduced as
\begin{equation}
\delta_{pq} = \delta_{n_{p}n_{q}} \delta_{l_{p}l_{q}} \delta_{j_{p}j_{q}} \delta_{t_{z,p}t_{z,q}}.
\end{equation}
Note that the factor $\sqrt{1/(1+\delta_{pq})}$ in the two-body state is for the normalization so that we have $\bra pq: J'M' | pq:JM \ket = \delta_{J'J} \delta_{M'M}$.
On the other hand, such normalization factor is not usually included in the three-body state.
For the three-body state, one can define another state by a different angular momentum coupling order, which should be related with the Wigner's $6j$-symbol.
In this paper, the first and second indices are always coupled first, and then the third index is coupled.
In practical applications, we need permutations of the indices of the states to further reduce the storage requirement.
For the two-body state, it is given by
\begin{equation}
|qp:JM \ket = -(-1)^{j_{p}+j_{q}-J} |pq:JM\ket.
\end{equation}
Likewise, permutations of the indices in the three-body state are given by
\begin{align}
\label{eq:qrp}
|qrp:J_{qr}JM\ket &= -\sum_{J_{pq}} (-1)^{j_{q}+j_{r}+J_{qr}} \sqrt{[J_{pq}][J_{qr}]}
\notag \\ &\hspace{-2em} \times
 \sixj{j_{p}}{j_{q}}{J_{pq}}{j_{r}}{J}{J_{qr}}|pqr:J_{pq}JM\ket, \\
 \label{eq:rpq}
 |rpq:J_{pr}JM\ket &= -\sum_{J_{pq}} (-1)^{j_{p}+j_{q}+J_{pq}} \sqrt{[J_{pq}][J_{pr}]}
\notag \\ &\hspace{-2em} \times
 \sixj{j_{p}}{j_{q}}{J_{pq}}{J}{j_{r}}{J_{pr}}|pqr:J_{pq}JM\ket, \\
  \label{eq:qpr}
|qpr:J_{pq}JM\ket &= -(-1)^{j_{p}+j_{q}-J_{pq}} |pqr:J_{pq}JM\ket, \\
\label{eq:rqp}
|rqp:J_{qr}JM\ket &= \sum_{J_{pq}}\sqrt{[J_{pq}][J_{qr}]}
\notag \\ &\hspace{-2em} \times
 \sixj{j_{p}}{j_{q}}{J_{pq}}{j_{r}}{J}{J_{qr}}|pqr:J_{pq}JM\ket, \\
 \label{eq:prq}
|prq:J_{pr}JM\ket &= -\sum_{J_{pq}} (-1)^{j_{q}+j_{r}+J_{pr}+J_{pq}} \sqrt{[J_{pq}][J_{pr}]}
\notag \\ &\hspace{-2em} \times
 \sixj{j_{p}}{j_{q}}{J_{pq}}{J}{j_{r}}{J_{pr}}|pqr:J_{pq}JM\ket.
\end{align}
Here, Wigner's $6j$-symbol with the standard notation~\cite{Varshalovich1988} and $[x]=2x+1$ are introduced.
With the $J$-coupled states, the two- and three-body matrix elements can be introduced as
$\bra p'q':J'M' | O^{[2]}| pq:JM \ket$ and $\bra p'q'r':J_{p'q'}J'M' | O^{[3]}| pqr:J_{pq}JM \ket$, respectively.
Because of the rotational invariance of the Hamiltonian, the Hamiltonian matrix is $M$-independent and diagonal with respect to $J$ and $M$.
Therefore, we introduce a shorthand notation for the matrix elements.
\begin{align}
T_{p'p} &= \bra p | T | q \ket, \\
T^{{\rm NN},J}_{p'q'pq} &= \bra p'q':JM | T^{\rm NN} | pq:JM \ket, \\
V^{{\rm NN},J}_{p'q'pq} &= \bra p'q':J'M' | V^{\rm NN}| pq:JM \ket, \\
V^{{\rm 3N},J_{p'q'}J_{pq}J}_{p'q'r'pqr} &= \bra p'q'r':J_{p'q'}J'M' | V^{\rm 3N} | pqr:J_{pq}JM \ket.
\end{align}
Since the uncoupled matrix elements with tilde indices can be computed from the $J$-coupled matrix elements, only calculating the $J$-coupled matrix elements is sufficient for many-body calculations.
The relation between the uncoupled and $J$-coupled matrix elements are the following.
\begin{align}
T_{\tilde{p}'\tilde{p}} &= T_{p'p} \delta_{m_{p'}m_{p}}, \\
T^{{\rm NN}}_{\tilde{p}'\tilde{q}'\tilde{p}\tilde{q}} &= \sum_{J} \cg{j_{p'}}{j_{q'}}{J}{m_{p'}}{m_{q'}}{M}
\cg{j_{p}}{j_{q}}{J}{m_{p}}{m_{q}}{M} T^{{\rm NN}, J}_{p'q'pq}, \\
V^{{\rm NN}}_{\tilde{p}'\tilde{q}'\tilde{p}\tilde{q}} &= \sum_{J} \cg{j_{p'}}{j_{q'}}{J}{m_{p'}}{m_{q'}}{M}
\cg{j_{p}}{j_{q}}{J}{m_{p}}{m_{q}}{M} V^{{\rm NN}, J}_{p'q'pq}, \\
V^{{\rm 3N}}_{\tilde{p}'\tilde{q}'\tilde{r}'\tilde{p}\tilde{q}\tilde{r}} &= \sum_{JJ_{p'q'}J_{pq}}
\cg{j_{p'}}{j_{q'}}{J_{p'q'}}{m_{p'}}{m_{q'}}{M_{p'q'}}
\cg{j_{p}}{j_{q}}{J_{pq}}{m_{p}}{m_{q}}{M_{pq}}
\notag \\ & \hspace{-2em} \times
\cg{J_{p'q'}}{j_{r'}}{J}{M_{p'q'}}{m_{r'}}{M}
\cg{J_{pq}}{j_{r}}{J}{M_{pq}}{m_{r}}{M}
V^{{\rm 3N}, J_{p'q'}J_{pq}J}_{p'q'r'pqr}.
\end{align}

\subsection{Matrix elements of kinetic terms}
The one-body kinetic matrix element $T_{p'p}$ is given by
\begin{equation}
\begin{aligned}
T_{p'p} &= \frac{A-1}{A} \delta_{l_{p'}l_{p}}\delta_{j_{p'}j_{p}} \delta_{t_{z,p'}t_{z,p}}
\frac{\hbar\omega}{2}
\\ & \times
\left[
\left(2n_{p} + l_{p} + \frac{3}{2}\right) \delta_{n_{p'}n_{p}}
\right. \\ & \left.  +
\sqrt{n_{p}\left(n_{p} + l_{p} + \frac{1}{2}\right)} \delta_{n_{p'}n_{p}-1}
\right. \\ & \left.  +
\sqrt{\left(n_{p}+1\right)\left(n_{p} + l_{p} + \frac{3}{2}\right)} \delta_{n_{p'}n_{p}+1}
\right].
\end{aligned}
\end{equation}
Note that the one-body kinetic operator takes the tridiagonal form.
Also, the two-body kinetic matrix element can be computed through the non-antisymmetrized $J$-coupled matrix element $\bar{T}^{{\rm NN}, J}_{p'q'pq}$
\begin{equation}
\begin{aligned}
\bar{T}^{{\rm NN}, J}_{p'q'pq} &=
 (-1)^{j_{q'}+j_{p}+J} \frac{\hbar^{2}}{Am}
 \\ & \times
\sixj{j_{p'}}{j_{q'}}{J}{j_{q}}{j_{p}}{1}
\bra p' \| \nabla \| p \ket
\bra q' \| \nabla \| q \ket,
\end{aligned}
\end{equation}
with the reduced matrix element of the gradient operator, which is given by
\begin{equation}
\begin{aligned}
 & \bra p' \| \nabla \| p \ket = (-1)^{l_{p}' + j_{p} + 1/2} \frac{1}{b} \sqrt{[j_{p'}][j_{p}]}
\sixj{j_{p'}}{j_{p}}{1}{l_{p}}{l_{p'}}{1/2}
\\ & \times
\left[
\sqrt{(l_{p}+1)(n_{p}+l_{p}+3/2)} \delta_{n_{p'}n_{p}}\delta_{l_{p'}l_{p}+1}
\right. \\ &\left.  \hspace{2em} +
\sqrt{(l_{p}+1)n_{p}} \delta_{n_{p'}n_{p}-1} \delta_{l_{p'}l_{p}+1}
\right. \\ &\left.  \hspace{2em}+
\sqrt{l_{p}(n_{p}+l_{p}+1/2)} \delta_{n_{p'}n_{p}} \delta_{l_{p'}l_{p}-1}
\right. \\ &\left. \hspace{2em} +
\sqrt{l_{p}(n_{p}+1)} \delta_{n_{p'}n_{p}+1} \delta_{l_{p'}l_{p}-1}
\right].
\end{aligned}
\end{equation}
Here, the HO length parameter $b^{2} \equiv \hbar / m\omega$ is introduced with the HO frequency $\omega$.
The antisymmetrized matrix element is obtained as
\begin{equation}
T^{{\rm NN}, J}_{p'q'pq} = \sqrt{\frac{1}{(1+\delta_{p'q'})(1+\delta_{pq})}} \left[
\bar{T}^{{\rm NN}, J}_{p'q'pq} - (-1)^{j_{p}+j_{q}-J}\bar{T}^{{\rm NN}, J}_{p'q'qp}\right],
\end{equation}
with
\begin{equation}
\delta_{pq} = \delta_{n_{p}n_{q}} \delta_{l_{p}l_{q}} \delta_{j_{p}j_{q}} \delta_{t_{z,p}t_{z,q}}.
\end{equation}
The main tasks remaining are to compute the matrix elements $V^{{\rm NN}, J}_{p'q'pq}$  and $V^{{\rm 3N},J_{p'q'}J_{pq}J}_{p'q'r'pqr}$.

\subsection{Nucleon-nucleon matrix elements}\label{sec:NN}
We begin with the NN matrix element.
One might think that the matrix element can be calculated directly from the integral using the single-particle HO wave function.
It is actually done in quantum chemistry.
However, this would be a computationally expensive task since functional forms of NN interactions are complicated.
Instead, the Talmi-Moshinsky transformation is widely used in nuclear physics:
\begin{equation}
\begin{aligned}
\label{eq:NN-trans}
V^{{\rm NN}, J}_{p'q'pq} &= \sum_{N^{\rm NN}_{\rm cm}L^{\rm NN}_{\rm cm}J^{\rm NN}_{\rm rel}S} \sum_{n'l'nl}
T^{p'q'J}_{N^{\rm NN}_{\rm cm}L^{\rm NN}_{\rm cm}n'l'SJ^{\rm NN}_{\rm rel}}
\\ & \hspace{2em}\times
V^{SJ^{\rm NN}_{\rm rel}}_{n'l'nl} \
T^{pqJ}_{N^{\rm NN}_{\rm cm}L^{\rm NN}_{\rm cm}nlSJ^{\rm NN}_{\rm rel}}.
\end{aligned}
\end{equation}
The quantum numbers introduced for the transformation $N^{\rm NN}_{\rm cm}$, $L^{\rm NN}_{\rm cm}$, $n$, $l$, $S$, and $J^{\rm NN}_{\rm rel}$ are the NN cm radial quantum number, NN cm orbital angular momentum, relative radial quantum number, relative orbital angular momentum, total spin, and total angular momentum of the relative motion, respectively.
The transformation coefficient $T^{pqJ}_{N^{\rm NN}_{\rm cm}L^{\rm NN}_{\rm cm}nlSJ^{\rm NN}_{\rm rel}}$ is
\begin{equation}
\begin{aligned}
\label{eq:tcoef-2b}
 T^{pqJ}_{N^{\rm NN}_{\rm cm}L^{\rm NN}_{\rm cm}nlSJ^{\rm NN}_{\rm rel}} &=
 (-1)^{L^{\rm NN}_{\rm cm}+l+S+J} \sqrt{[j_{p}][j_{q}][S][J^{\rm NN}_{\rm rel}]}
 \\  & \hspace{-2em}\times
 \sum_{\Lambda} [\Lambda]
\ninej{l_{p}}{1/2}{j_{p}}{l_{q}}{1/2}{j_{q}}{\Lambda}{S}{J}
\sixj{L^{\rm NN}_{\rm cm}}{l}{\Lambda}{S}{J}{J^{\rm NN}_{\rm rel}}
 \\ &  \hspace{-2em} \times
\bra N^{\rm NN}_{\rm cm}L^{\rm NN}_{\rm cm} nl: \Lambda \vert n_{p}l_{p}n_{q}l_{q} : \Lambda \ket_{1}.
\end{aligned}
\end{equation}
In the above equation, 9$j$-symbol is used with the standard notation~\cite{Varshalovich1988}.
The symbol $\bra NLnl:\Lambda \vert n_{1}l_{1}n_{2}l_{2} : \Lambda \ket_{d}$ is the HO bracket defined with the notation in Ref.~\cite{Kamuntavicius2001}.
The inner summations in Eq.~\eqref{eq:NN-trans} can be performed with an efficient matrix multiplication.
Note that the antisymmetrization is not taken into account here.
However, it is trivial and can be done by multiplying the factor $f_{pq}$ to Eq.~\eqref{eq:tcoef-2b}:
\begin{equation}
f_{pq} = \left\{
\begin{array}{cc}
1, & t_{z,p} \neq t_{z,q} \\
\sqrt{\frac{1}{2(1+\delta_{pq})}} [1 + (-1)^{l + S}], & t_{z,p} = t_{z,q}
\end{array}
\right. .
\end{equation}
The NN matrix element in the relative HO basis $V^{SJ^{\rm NN}_{\rm rel}}_{n'l'nl}$ can be obtained through the integral:
\begin{equation}
\begin{aligned}
V^{SJ^{\rm NN}_{\rm rel}}_{n'l'nl} &= \int d\pi'_{1} d\pi_{1} \ \pi'^{2}_{1} \pi^{2}_{1} R_{n'l'}(\pi'_{1}) R_{nl}(\pi_{1})
\\ & \times
V^{SJ^{\rm NN}_{\rm rel}}_{l'l} (\pi'_{1},\pi_{1}),
\end{aligned}
\end{equation}
with the radial HO wave function:
\begin{equation}
\begin{aligned}
R_{nl}(\pi_{1}) &= (-1)^{n} b \sqrt{\frac{2b\Gamma(n+1)}{\Gamma(n+l+3/2)}}
\\ & \times (\pi_{1}b)^{l}
e^{-\pi^{2}_{1}b^{2}/2} L^{(l+1/2)}_{n}(\pi^{2}_{1}b^{2}).
\end{aligned}
\end{equation}
The gamma function $\Gamma(x)$ and associated Laguerre polynomial $L^{(\alpha)}_{n}(x)$ are introduced.
The momentum $\pi_{1}$ is $\pi_{1} = \vert(\vec{p}_{1}-\vec{p}_{2})\vert /\sqrt{2}$, consistent with the definition of the HO bracket.
Note that $\pi_{1}$ is different from the usual relative momentum definition $p = \vert \vec{p}_{1} - \vec{p}_{2}\vert /2$.

The \nuhamil code requires the input file for $V^{SJ^{\rm NN}_{\rm rel}}_{l'l} (p',p)$ stored as a function of $p'$ and $p$.
Some selected interactions are given in the \texttt{input\_nn\_files} directory.
The available NN interactions are
LO -- N$^{4}$LO with 500 MeV regulator cutoff by Entem--Machleidt--Nosyk~\cite{Entem2017}, N$^{3}$LO with 500 MeV regulator cutoff by Entem--Machleidt~\cite{Entem2003}, N$^{2}$LO$_{\rm opt}$~\cite{Ekstrom2013}, N$^{2}$LO$_{\rm sat}$~\cite{Ekstrom2015}, and $\Delta$-full EFT series by the Gothenburg--Oak Ridge collaboration~\cite{Jiang2020}.

\subsection{Three-nucleon matrix elements}\label{sec:3N}
For computational reasons, the 3N matrix elements are calculated within the isospin formalism.
The matrix elements with the proton-neutron basis can be obtained through the $JT$-coupled matrix element $V^{{\rm 3N}, J_{p'q'}J_{pq}J, T_{p'q'}T_{pq}T}_{p'q'r'pqr}$:
\begin{equation}
\begin{aligned}
V^{{\rm 3N}, J_{\rm p'q'}J_{pq}J}_{p'q'r'pqr} &=
\sum_{T_{p'q'}T_{pq}T}
\cg{t_{p'}}{t_{q'}}{T_{p'q'}}{t_{z,p'}}{t_{z,q'}}{T_{z,p'q'}}
\cg{t_{p}}{t_{q}}{T_{pq}}{t_{z,p}}{t_{z,q}}{T_{z,pq}}
\\ &\times
\cg{T_{p'q'}}{t_{r'}}{T}{T_{z,p'q'}}{t_{z,r'}}{T_{z}}
\cg{T_{pq}}{t_{r}}{T}{T_{z,pq}}{t_{z,r}}{T_{z}}
V^{{\rm 3N}, J_{p'q'}J_{pq}J, T_{p'q'}T_{pq}T}_{p'q'r'pqr}.
\end{aligned}
\end{equation}
The recoupling should be done in many-body calculations, and the goal here is to obtain the $JT$-coupled matrix element.
Note that recoupling coefficients from isopin structure have to be considered for the permutation of indices, similar to Eqs.~\eqref{eq:qrp}-\eqref{eq:prq}.

Since the antisymmetrization of the 3N basis is more complicated than that of the NN basis, the 3N matrix element is cumbersome.
The antisymmetrized basis is expressed as the linear combination of the non-antisymmetrized basis:
\begin{equation}
 \vert EiJ^{3N}_{\rm rel}T \ket = \sum_{\beta} c_{i\beta} \vert E\beta J^{3N}_{\rm rel}T \ket,
\end{equation}
where $E$, $i$, and $J^{3N}_{\rm rel}$ are the HO principle quantum number, label distinguishing the states, and total Jacobi angular momentum, respectively.
The collective index $\beta$,
\begin{equation}
\beta = \{n_{12}, l_{12}, s_{12}, j_{12}, t_{12}, n_{3}, l_{3}, j_{3}\},
\end{equation}
specifies the non-antisymmetrized basis. The quantum numbers with the subscript `12', $n_{12}$, $l_{12}$, $s_{12}$, $j_{12}$, and $t_{12}$ are for the relative motion of nucleons 1 and 2, i.e., the nodal, orbital angular momentum, spin, total angular momentum, and total isospin quantum numbers, respectively.
Likewise, $n_{3}$, $l_{3}$, and $j_{3}$ are the quantum numbers for the nucleon 3 with respect to the cm of the nucleons 1 and 2.
Note that the principle quantum number is defined as $E=2n_{12}+l_{12}+2n_{3}+l_{3}$.
The coefficient in the linear combination $c_{i\beta}$ can be obtained by the diagonalization of the antisymmetrizer~\cite{Navratil1999,Navratil2000}:
\begin{equation}
\mathcal{A} \vert EiJ^{3N}_{\rm rel}T \ket = A_{i} \vert EiJ^{3N}_{\rm rel}T \ket,
\end{equation}
with $\mathcal{A} = (1 + \mathcal{T}_{13}\mathcal{T}_{12} + \mathcal{T}_{12}\mathcal{T}_{23}-\mathcal{T}_{12}-\mathcal{T}_{13}-\mathcal{T}_{23})/6$ defined with the exchange operator $\mathcal{T}_{ij}$ and the eigenvalue $A_{i}$.
The matrix element of the antisymmetrizer is~\cite{Navratil1999,Navratil2000}
\begin{equation}
\begin{aligned}
\bra & E'\beta'  J^{{\rm 3N}'}_{\rm rel}T' \vert \mathcal{A} \vert E\beta J^{{\rm 3N}}_{\rm rel}T \ket =
\\ & \times \left[ \frac{\delta_{\beta'\beta}}{3} -
\frac{2}{3} (-1)^{s'_{12}+t'_{12}+s_{12}+t_{12}} \sum_{\Lambda S} [\Lambda][S]
\right. \\ & \left. \times
\sqrt{[s'_{12}][j'_{12}][j'_{3}][t'_{12}]}
\sqrt{[s_{12}][j_{12}][j_{3}][t_{12}]}
\right. \\ & \left. \times
\ninej{l'_{12}}{s'_{12}}{j'_{12}}{l'_{3}}{1/2}{j'_{3}}{\Lambda}{S}{J}
\ninej{l_{12}}{s_{12}}{j_{12}}{l_{3}}{1/2}{j_{3}}{\Lambda}{S}{J}
\right. \\ & \left. \times
\sixj{1/2}{1/2}{s'_{12}}{1/2}{S}{s_{12}}
\sixj{1/2}{1/2}{t'_{12}}{1/2}{T}{t_{12}}
\right. \\ & \left. \times
\bra n'_{12}l'_{12}n'_{3}l'_{3}: \Lambda \vert n_{12}l_{12}n_{3}l_{3} \Lambda \ket_{1/3}
\right]
\\ & \times
\delta_{E'E}\delta_{J^{{\rm 3N}'}_{\rm rel}J^{{\rm 3N}}_{\rm rel}} \delta_{T'T}.
\end{aligned}
\end{equation}
The eigenvalue problem can be separated into $\{E, J^{{\rm 3N}}_{\rm rel}, T\}$ blocks.
Due to the overcompleteness of the non-antisymmetrized basis, the eigenvalue $A_{i}$ is either $0$ or $1$, and $A_{i}$ for the physical state has to be $1$.
Therefore, we always see $N_{\rm A} \leq N_{\rm NA}$ where $N_{\rm A}$ and $N_{\rm NA}$ are the basis numbers in the antisymmetrized and non-antisymmetrized bases within the $\{E, J^{{\rm 3N}}_{\rm rel}, T\}$ block, respectively.
In the code, all the 3N operators are stored with the antisymmetrized basis rather than the non-antisymmetrized basis, as it is computationally easier to handle.
As another option for the 3N antisymmetrization, one may apply the permutator operator to the 3N momentum state as introduced in Ref.~\cite{Glockle1983}.

Similarly to the NN case, the 3N matrix elements can be obtained through the three-body Talmi-Moshinsky transformation:
\begin{equation}
\begin{aligned}
\label{eq:3N-trans}
& V^{{\rm 3N}, J_{ p'q'}J_{pq}J, T_{p'q'}T_{pq}T}_{p'q'r'pqr} = 6 \sum_{N^{\rm 3N}_{\rm cm}L^{\rm 3N}_{\rm cm}J^{\rm 3N}_{\rm rel}} \sum_{E'i'Ei}
\\ & \times
T^{p'q'r'J_{p'q'}J T_{p'q'}T}_{N^{\rm 3N}_{\rm cm}L^{\rm 3N}_{\rm cm} E'i'J^{\rm 3N}_{\rm rel}} \ V^{J^{\rm 3N}_{\rm rel}T}_{E'i'Ei}
T^{pqrJ_{pq}JT_{pq}T}_{N^{\rm 3N}_{\rm cm}L^{\rm 3N}_{\rm cm}Ei J^{\rm 3N}_{\rm rel}},
\end{aligned}
\end{equation}
where $N^{\rm 3N}_{\rm cm}$ and $L^{\rm 3N}_{\rm cm}$ denote the 3N cm radial quantum number and 3N cm orbital angular momentum, respectively.
The matrix element $V^{J^{\rm 3N}_{\rm rel}}_{E'i'Ei}$ is a shorthand notation of $\bra E'i'J^{\rm 3N}_{\rm rel}T \vert V \vert EiJ^{\rm 3N}_{\rm rel}T\ket$.
The transformation coefficient is given by
\begin{equation}
T^{pqr J_{pq}J T_{pq}T}_{N^{\rm 3N}_{\rm cm}L^{\rm 3N}_{\rm cm}Ei J^{\rm 3N}_{\rm rel}} = \sum_{\beta} c_{i\beta} T^{pqr J_{pq}J T_{pq}T}_{N^{\rm 3N}_{\rm cm}L^{\rm 3N}_{\rm cm} E\beta J^{\rm 3N}_{\rm rel}},
\end{equation}
with
\begin{equation}
\begin{aligned}
& T^{pqrJ_{pq}JT_{pq}T}_{N^{\rm 3N}_{\rm cm}L^{\rm 3N}_{\rm cm}E\beta J^{\rm 3N}_{\rm rel}}  = \delta_{T_{pq}t_{12}} (-1)^{L^{\rm 3N}_{\rm cm}+j_{3}+3/2}
\sqrt{[J_{pq}][j_{r}][J^{\rm 3N}_{\rm rel}][j_{3}]}
\\ & \times
\sum_{N^{\rm NN}_{\rm cm}L^{\rm NN}_{\rm cm}}
T^{pqJ_{pq}}_{N^{\rm NN}_{\rm cm}L^{\rm NN}_{\rm cm}n_{12}l_{12}s_{12}j_{12}}
\sum_{\Lambda} (-1)^{\Lambda} [\Lambda]
\\ &  \times
\bra N^{\rm 3N}_{\rm cm}L^{\rm 3N}_{\rm cm} n_{3}l_{3}: \Lambda \vert N^{\rm NN}_{\rm cm}L^{\rm NN}_{\rm cm} n_{r}l_{r} : \Lambda \ket_{2}
\\ & \times
\twlvj{j_{12}}{L^{\rm NN}_{\rm cm}}{\Lambda}{L^{\rm 3N}_{\rm cm}}{J_{pq}}{l_{r}}{l_{3}}{J^{\rm 3N}_{\rm rel}}{J}{j_{r}}{1/2}{j_{3}}.
\end{aligned}
\end{equation}
The 12$j$-symbol of the first kind~\cite{Varshalovich1988} is used.
Note that $T^{pqJ_{pq}}_{N^{\rm NN}_{\rm cm}L^{\rm NN}_{\rm cm}n_{12}l_{12}s_{12}j_{12}}$ is defined in Eq.~\eqref{eq:tcoef-2b}, and one can find a recursive relation for the $N$-body Talmi-Moshinsky transformation with the 12$j$-symbol and HO bracket.

A typical limit of the three-body Talmi-Moshinky transformation is $E_{\rm 3max}=16$, where $E_{\rm 3max}$ is defined as $\max(2n_{p}+l_{p}+2n_{q}+l_{q}+2n_{r}+l_{r})$.
This limit does not allow us to obtain converged results for heavier systems with $A \gtrsim 100$.
Recently, the limit was extended to $E_{\rm 3max}=28$~\cite{Miyagi2022}, by only computing the matrix elements relevant to the normal-ordered two-body (NO2B) approximation, which is widely used in basis expansion methods.
For further details about the NO2B matrix elements, see Ref.~\cite{Miyagi2022}.
The code supports this format as well.

The 3N matrix element in the Jacobi HO basis $V^{J^{\rm 3N}_{\rm rel}T}_{E'i'Ei}$ can be obtained from the matrix element expressed with the non-antisymmetrized basis:
\begin{equation}
V^{J^{\rm 3N}_{\rm rel}T}_{E'i'Ei} = \sum_{\beta'\beta} c_{i'\beta'} c_{i\beta} V^{J^{\rm 3N}_{\rm rel}T}_{\beta'\beta},
\end{equation}
with
\begin{equation}
\begin{aligned}
\label{eq:mom_to_ho_3b}
& V^{J^{\rm 3N}_{\rm rel}T}_{\beta'\beta} = \int  d\pi_{1}' d\pi_{2}' d\pi_{1} d\pi_{2} \ \pi'^{2}_{1} \pi'^{2}_{2} \pi^{2}_{1} \pi^{2}_{2}
\\ & \times
R_{n'_{12}l'_{12}}(\pi'_{1}) R_{n_{12}l_{12}}(\pi_{1})
R_{n'_{3}l'_{3}}(\pi'_{2}) R_{n_{3}l_{3}}(\pi_{2})
\\ & \times
V^{J^{\rm 3N}_{\rm rel}T}_{\alpha'\alpha}(\pi'_{1},\pi'_{2},\pi_{1},\pi_{2}).
\end{aligned}
\end{equation}
Here, the collective index $\alpha$ is introduced as
\begin{equation}
\label{eq:def_alpha}
\alpha = \{l_{12},s_{12}, j_{12}, t_{12}, l_{3}, j_{3}\}.
\end{equation}
The momentum $\pi_{2}$ is defined as $\pi_{2} = \sqrt{2/3} | (\vec{p}_{1}+\vec{p}_{2})/2 - \vec{p}_{3} |$.
The momentum-space matrix element includes a regulator function:
\begin{equation}
V^{J^{\rm 3N}_{\rm rel}T}_{\alpha'\alpha}(\pi'_{1},\pi'_{2},\pi_{1},\pi_{2}) = f_{\Lambda}
V^{J^{\rm 3N}_{\rm rel}T}_{\chi{\rm EFT}, \alpha'\alpha}(\pi'_{1},\pi'_{2},\pi_{1},\pi_{2}).
\end{equation}
The regulator function takes either a non-local form
\begin{equation}
\label{eq:freg_nonlocal}
f_{\Lambda_{\rm nonlocal}} = \exp\left[-\left(\frac{\pi'^{2}_{1} + \pi'^{2}_{2}}{2\Lambda^{2}_{\rm nonlocal}}\right)^{n}\right]
\exp\left[-\left(\frac{\pi^{2}_{1} + \pi^{2}_{2}}{2\Lambda^{2}_{\rm nonlocal}}\right)^{n}\right],
\end{equation}
a local form
\begin{equation}
\label{eq:freg_local}
f_{\Lambda_{\rm local}} = \exp\left[-\left(\frac{|\vec{p}'_{1} - \vec{p}_{1}|^{2}}{\Lambda^{2}_{\rm local}}\right)^{n}\right]
\exp\left[-\left(\frac{|\vec{p}'_{2} - \vec{p}_{2}|^{2}}{\Lambda^{2}_{\rm local}}\right)^{n}\right],
\end{equation}
or a semilocal form~\cite{Epelbaum2020} that is not supported in the code.
The code fully supports the locally regulated matrix elements at N$^{2}$LO in chiral EFT based on Ref.~\cite{Navratil2007}.
Also, a newly introduced local-non-local regularized form, $f_{\Lambda}=f_{\Lambda_{\rm local}}f_{\Lambda_{\rm nonlocal}}$~\cite{Soma2020}, is supported.

For non-local matrix elements, we have tried to implement along Ref.~\cite{Fukui2018}.
However, we found a numerical instability in the higher angular momentum partial waves, which would be related to the discussion made in Ref.~\cite{Huber1997}.
For this reason, non-local matrix elements are not fully supported, and external input files are required.
The code has the capability to read the momentum-space matrix element $V^{J^{\rm 3N}_{\rm rel}T}_{\chi {\rm EFT},\alpha'\alpha}(p',q',p,q)$\footnote{In the HDF files, the matrix elements are given in terms of the Jacobi variables $p$ and $q$~\cite{Hebeler2021}, defined as
$$
p = \frac{1}{2} |\vec{p}_{1} - \vec{p}_{2}| = \sqrt{\frac{1}{2}} \pi_{1},
$$
and
$$
q = \frac{1}{3} | -\vec{p}_{1} - \vec{p}_{2} + 2\vec{p}_{3}| = \sqrt{\frac{2}{3}}|-\vec{\pi}_{2}|.
$$
} from the HDF files by Hebeler et al.~\cite{Hebeler2015,Hebeler2021}.
In the integral~\eqref{eq:mom_to_ho_3b}, the cubic b-spline interpolation is used to capture the oscillating nature of the HO wave functions.

\begin{table*}
 \caption{The ground-state energies of selected doubly magic nuclei computed with the NCSM and IMSRG(2). The resolution scale $\lambda$ is related with the end point of the SRG evolution: $\lambda = \alpha^{-1/4}$ in units of fm$^{-1}$. The $N_{\rm max}$ and $e_{\rm max}$ truncations are employed in the NCSM and IMSRG(2), respectively.
 The truncation defined by $E_{\rm 3max} = \max(e_{1}+e_{2}+e_{3})$ is applied for the input 3N matrix elements. The $E_{\rm 3max}$ values with asterisk indicates that the 3N matrix elements are computed within the half-precision floating-point numbers.
 Also, $E_{\rm 3max}$=``none'' indicates that there are no input 3N matrix elements.
 The entry $\hbar\omega$ is the basis frequency used in the many-body calculations, and the numbers in the parentheses are the parent frequency adopted in the frequency conversion for the 3N matrix elements~\cite{Roth2014}. For $^{4}$He, $^{16}$O, and $^{40}$Ca, the 3N SRG evolution was done in the ramp A space defined in Ref.~\cite{Roth2014}. For $^{132}$Sn, the evolution was done in the $N_{\rm max}=48$ space only for $J^{3N}_{\rm rel} \leq 13/2$ channels. The star symbol in the $E_{\rm 3max}$ column indicates that the 3N matrix elements are in half-precision floating point numbers.}
  \label{tab:res}
\begin{tabular}{ c c c c c c c c } \hline \hline
 Nucleus & Interaction & $\lambda$ (fm$^{-1}$) & Method & $N_{\rm max}$/$e_{\rm max}$ & $E_{\rm 3max}$ & $\hbar\omega$ (MeV) & $E_{\rm g.s.}$ (MeV)   \\ \hline
 $^{4}$He & EM500 \cite{Entem2003} & 1.8 & NCSM & 16 & none & 16 & $-28.45$ \\
 $^{4}$He & EM500 \cite{Entem2003} & 2.0 & NCSM & 16 & none & 16 & $-28.23$ \\
 $^{4}$He & EM500 \cite{Entem2003} & 2.2 & NCSM & 16 & none & 16 & $-27.88$ \\
 $^{4}$He & EM500 \cite{Entem2003} & 1.8 & IMSRG(2) & 14 & none & 16 & $-28.43$ \\
 $^{4}$He & EM500 \cite{Entem2003} & 2.0 & IMSRG(2) & 14 & none & 16 & $-28.28$ \\
 $^{4}$He & EM500 \cite{Entem2003} & 2.2 & IMSRG(2) & 14 & none & 16 & $-28.08$ \\
 $^{4}$He & EM500 \cite{Entem2003} & 1.8 & IMSRG(2) & 14 & 16 & 16(30) & $-25.61$ \\
 $^{4}$He & EM500 \cite{Entem2003} & 2.0 & IMSRG(2) & 14 & 16 & 16(30) & $-25.76$ \\
 $^{4}$He & EM500 \cite{Entem2003} & 2.2 & IMSRG(2) & 14 & 16 & 16(30) & $-25.96$ \\
 $^{4}$He & N$^{3}$LO$_{\rm lnl}$ \cite{Soma2020} & 1.8 & IMSRG(2) & 14 & 16 & 16(30) & $-28.63$ \\
 $^{4}$He & N$^{3}$LO$_{\rm lnl}$ \cite{Soma2020} & 2.0 & IMSRG(2) & 14 & 16 & 16(30) & $-28.69$ \\
 $^{4}$He & N$^{3}$LO$_{\rm lnl}$ \cite{Soma2020} & 2.2 & IMSRG(2) & 14 & 16 & 16(30) & $-28.76$ \\
 $^{4}$He & 1.8/2.0 (EM) \cite{Hebeler2011} & none & IMSRG(2) & 14 & 16 & 16 & $-29.23$ \\
 $^{4}$He & $\Delta$N$^{2}$LO$_{\rm GO}$(394) \cite{Jiang2020} & none & IMSRG(2) & 14 & 16 & 16 & $-28.67$ \\
 & \\
 $^{16}$O & EM500 \cite{Entem2003} & 1.8 & IMSRG(2) & 14 & none & 16 & $-173.3$ \\
 $^{16}$O & EM500 \cite{Entem2003} & 2.0 & IMSRG(2) & 14 & none & 16 & $-165.7$ \\
 $^{16}$O & EM500 \cite{Entem2003} & 2.2 & IMSRG(2) & 14 & none & 16 & $-158.5$ \\
 $^{16}$O & EM500 \cite{Entem2003} & 1.8 & IMSRG(2) & 14 & 16 & 16(30) & $-121.6$ \\
 $^{16}$O & EM500 \cite{Entem2003} & 2.0 & IMSRG(2) & 14 & 16 & 16(30) & $-122.9$  \\
 $^{16}$O & EM500 \cite{Entem2003} & 2.2 & IMSRG(2) & 14 & 16 & 16(30) & $-124.2$  \\
 $^{16}$O & N$^{3}$LO$_{\rm lnl}$ \cite{Soma2020} & 1.8 & IMSRG(2) & 14 & 16 & 16(30) & $-128.6$ \\
 $^{16}$O & N$^{3}$LO$_{\rm lnl}$ \cite{Soma2020} & 2.0 & IMSRG(2) & 14 & 16 & 16(30) & $-127.2$ \\
 $^{16}$O & N$^{3}$LO$_{\rm lnl}$ \cite{Soma2020} & 2.2 & IMSRG(2) & 14 & 16 & 16(30) & $-126.0$ \\
 $^{16}$O & 1.8/2.0 (EM) \cite{Hebeler2011} & none & IMSRG(2) & 14 & 16 & 16 & $-127.2$ \\
 $^{16}$O & $\Delta$N$^{2}$LO$_{\rm GO}$(394) \cite{Jiang2020} & none & IMSRG(2) & 14 & 16 & 16 & $-126.1$ \\  & \\
 $^{40}$Ca & EM500 \cite{Entem2003} & 1.8 & IMSRG(2) & 14 & none & 16 & $-639.4$ \\
 $^{40}$Ca & EM500 \cite{Entem2003} & 2.0 & IMSRG(2) & 14 & none & 16 & $-595.8$ \\
 $^{40}$Ca & EM500 \cite{Entem2003} & 2.2 & IMSRG(2) & 14 & none & 16 & $-554.1$ \\
 $^{40}$Ca & EM500 \cite{Entem2003} & 1.8 & IMSRG(2) & 14 & 16 & 16(30) & $-352.3$ \\
 $^{40}$Ca & EM500 \cite{Entem2003} & 2.0 & IMSRG(2) & 14 & 16 & 16(30) & $-360.0$ \\
 $^{40}$Ca & EM500 \cite{Entem2003} & 2.2 & IMSRG(2) & 14 & 16 & 16(30) & $-366.7$  \\
 $^{40}$Ca & N$^{3}$LO$_{\rm lnl}$ \cite{Soma2020} & 1.8 & IMSRG(2) & 14 & 16 & 16(30) & $-347.1$ \\
 $^{40}$Ca & N$^{3}$LO$_{\rm lnl}$ \cite{Soma2020} & 2.0 & IMSRG(2) & 14 & 16 & 16(30) & $-341.9$ \\
 $^{40}$Ca & N$^{3}$LO$_{\rm lnl}$ \cite{Soma2020} & 2.2 & IMSRG(2) & 14 & 16 & 16(30) & $-336.7$ \\
 $^{40}$Ca & 1.8/2.0 (EM) \cite{Hebeler2011} & none & IMSRG(2) & 14 & 16 & 16 & $-344.4$ \\
 $^{40}$Ca & $\Delta$N$^{2}$LO$_{\rm GO}$(394) \cite{Jiang2020} & none & IMSRG(2) & 14 & 16 & 16 & $-339.1$ \\ & \\
 $^{132}$Sn & N$^{3}$LO$_{\rm lnl}$ \cite{Soma2020} & 2.0 & IMSRG(2) & 14 & 24 & 12 (30) & $-1064$ \\
 $^{132}$Sn & 1.8/2.0 (EM) \cite{Hebeler2011} & none & IMSRG(2) & 14 & 24 & 12 & $-1109$  \\
 $^{132}$Sn & $\Delta$N$^{2}$LO$_{\rm GO}$(394) \cite{Jiang2020} & none & IMSRG(2) & 14 & 24 & 12 & $-1098$ \\ & \\
 $^{208}$Pb & 1.8/2.0 (EM) \cite{Hebeler2011} & none & IMSRG(2) & 14 & 28* & 12 & $-1660$ \\
 $^{208}$Pb & $\Delta$N$^{2}$LO$_{\rm GO}$(394) \cite{Jiang2020} & none & IMSRG(2) & 14 & 28* & 12 & $-1628$ \\
 \hline \hline
\end{tabular}
\end{table*}

\section{Similarity renormalization group}\label{sec:SRG}
The momentum scales of chiral EFT interactions are significantly lower than those of the other potential models such as AV18~\cite{Wiringa1995} and CD-Bonn~\cite{Machleidt2001}.
However, the momentum scale is not sufficiently low to obtain converged results in the many-body calculations.
To accelerate convergence, one sometimes softens nuclear interactions.
Softening procedures are well summarized for example in Ref.~\cite{Bogner2010}.
Here, we briefly review the widely used similarity renormalization group (SRG) approach.

In the SRG, we consider a unitary transformation depending on a continuous parameter $\alpha$:
\begin{equation}
H(\alpha) = U^{\dag}(\alpha)HU(\alpha).
\end{equation}
The SRG flow equation can be obtained by differentiating both sides:
\begin{equation}
\label{eq:flow}
\begin{aligned}
\frac{dH(\alpha)}{d\alpha} &= [\eta(\alpha), H(\alpha)], \quad
\eta(\alpha) = \frac{dU^{\dag}(\alpha)}{d\alpha} U(\alpha).
\end{aligned}
\end{equation}
The antihermitian operator $\eta(\alpha)$ is known as the generator of the flow equation and can be chosen flexibly~\cite{Wegner1994}.
The most widely used choice is $\eta(\alpha) = [T_{\rm kin}, H(\alpha)]$ with the kinetic operator $T_{\rm kin}$, which guarantees the suppression of the coupling between low and high momenta.
The flow equation~\eqref{eq:flow} is integrated until the coupling is sufficiently suppressed.
Since the unitary transformation does not change the eigenvalues, the SRG can be regarded as a reshuffling of the NN, 3N, and many-body sectors.
In other words, the many-body interactions are induced by the SRG evolution even if the original interaction includes only NN interactions.

In practical applications, we extract the SRG-evolved interactions with the subtraction method (see Ref.~\cite{Roth2014} for example).
For NN, the Hamiltonian $H_{\rm NN} = T_{\rm NN} + V_{\rm NN}$ is evolved, and the evolved NN interaction is given as $V_{\rm NN}(\alpha) = H_{\rm NN}(\alpha) - T_{\rm kin}$.
Note that $T_{\rm kin}$ and $V_{\rm NN}$ are the NN kinetic and interaction operators, respectively.
From the definition of the flow equation, $V_{\rm NN}(\alpha)$ reproduces the two-body observables obtained through the original interaction $V_{\rm NN}$.
However, this is not true for three- and many-body observables due to missing induced many-body forces, and the many-body observables have an artificial $\alpha$ dependence, showing how much the unitarity of the transformation is broken in the many-body space.
To obtain more $\alpha$-independent result, one has to include induced 3N interaction extracted from the 3N evolution.

For 3N, the starting Hamiltonian is $H_{\rm 3N} = T_{\rm 3N} + V^{[3]}_{\rm NN}$, where $T_{\rm 3N}$ is the 3N kinetic operator, and $V_{\rm NN}^{[3]}$ is the NN interaction embedded into the 3N space.
Note that one can add an initial 3N term if required.
The induced 3N term can be obtained as $V_{\rm 3N, ind}(\alpha) = H_{\rm 3N}(\alpha) - T_{\rm 3N} - V_{\rm NN}^{[3]}(\alpha)$.
Here, $V_{\rm NN}^{[3]}(\alpha)$ is $V_{\rm NN}(\alpha)$ obtained from the NN evolution and embedded into 3N space.
As seen in the NN evolution, the 3N evolution preserves three-body observables.
The same procedure can be applied for many-body terms, and the many-body evolution is needed until the $\alpha$-dependence of the many-body observables becomes weak enough.
However, in practice, even the four-body SRG evolution is too expensive to do due to the resulting basis dimension and the cost of antisymmetrizing the basis.
In the \nuhamil code, the SRG evolution can be performed in the NN and 3N sectors, the current state-of-the-art.

The unitary transformation can be obtained from the flow equation for the transformation operator:
\begin{equation}
\frac{dU(\alpha)}{d\alpha} = - U(\alpha) \eta(\alpha).
\end{equation}
However, a computationally more moderate way is used in practice, and
the unitary transformation is obtained as
\begin{equation}
U(\alpha) = \sum_{k} \vert \psi_{k} \ket \bra \psi_{k}(\alpha) \vert,
\end{equation}
with the eigenstates of the original and evolved Hamiltonians:
\begin{equation}
\begin{aligned}
H\vert \psi_{k} \ket = E_{k} \vert \psi_{k} \ket, \quad
H(\alpha)\vert \psi_{k}(\alpha) \ket = E_{k} \vert \psi_{k}(\alpha) \ket.
\end{aligned}
\end{equation}
Note that the relative phase of $\vert\psi_{k}\ket$ and $\vert\psi_{k}(\alpha)\ket$ cannot be determined in general, which affect the sign of the matrix element of the transformation operator.
Since the SRG transformation does not change the wave function drastically, the relative phase is fixed such that $\bra \psi_{k} \vert \psi_{k}(\alpha) \ket \geq 0$.
In the same way as for the Hamiltonian, the induced three-body term of an operator can be computed in the \nuhamil code as done in Ref.~\cite{Miyagi2019}.
The end point of the flow equation is usually parametrized by the momentum scale $\lambda = \alpha^{-1/4}$ instead of $\alpha$, and we follow this convention.

In the code, the SRG evolution is done in a relative-coordinate HO space because the consistent evolution of the other operators is straightforward.
This means that the evolution is done in the truncated HO space, and the $N_{\rm max}$ truncation is employed in the code.
The $N_{\rm max}$ is defined as
\begin{equation}
N_{\rm max} = \left\{
\begin{array}{cc}
\max(2n+l), & \text{NN system} \\
\max(2n_{12}+l_{12}+2n_{3}+l_{3}), & \text{3N system}
\end{array}
\right. .
\end{equation}
The UV momentum scale in the employed $N_{\rm max}$ space is roughly estimated as $p_{\rm UV} \sim \sqrt{2 N_{\rm max}m\omega}$~\cite{Furnstahl2012}, and we expect that the $N_{\rm max}$ should be increased until $p_{\rm UV}$ is sufficiently larger than the cutoff scale of the interaction, typically $500$ MeV.
Although we can take sufficiently large $N_{\rm max}$ for the NN evolution\footnote{We observed that HO- and momentum-space evolutions provide almost the same results. The code supports the momentum-space evolution only for NN interactions, and users can verify it.}, the 3N $N_{\rm max}$ can be an issue especially in heavy nuclei calculations~\cite{Miyagi2022} even if the frequency conversion technique~\cite{Roth2014} is used.

\section{Many-body results}\label{sec:Results}
Here, we show the ground-state energies for the selected doubly magic nuclei computed with the NCSM and two-body approximated IMSRG [IMSRG(2)] as a benchmark.
We do not introduce the theoretical details of the many-body calculation methods.
The details can be found in Refs.~\cite{Barrett2013,Hergert2016,Stroberg2019} and references therein.
The numerical codes used here are open source; the NCSM and IMSRG are done with the \texttt{BIGSTICK}~\cite{Johnson2018} and \texttt{imsrg++}~\cite{imsrg++} codes, respectively.

In Table~\ref{tab:res}, the ground-state energies are shown, using the SRG-softened N$^{3}$LO NN interaction~\cite{Entem2003}, labeled by ``EM500",  with and without induced 3N interaction, N$^{3}$LO$_{\rm lnl}$~\cite{Soma2020}, 1.8/2.0~(EM)~\cite{Hebeler2011}, and $\Delta$N$^{2}$LO$_{\rm GO}$(394)~\cite{Jiang2020}.

\section{Program Summary and Specifications\label{sec:Usage}}
The \nuhamil code is written in modern Fortran.
It requires a set of libraries, \texttt{BLAS}, \texttt{LAPACK}, GNU scientific library (\texttt{gsl}), \texttt{zlib}, and \texttt{hdf5}.

\subsection{Installation}
The source code can downloaded from GitHub:
\begin{minted}[breaklines,breakanywhere]{text}
 $ cd ~
 $ git clone https://github.com/Takayuki-Miyagi/NuHamil-public.git
\end{minted}
Note that downloading the code in the home directory is not mandatory, but it is recommended.
One needs to download the submodules, linear algebra wrapper and b-spline interpolation:
\begin{minted}{text}
 $ cd NuHamil
 $ git submodule init
 $ git submodule update
\end{minted}
The compilation can be done with the \texttt{make} command.
The default compiler is \texttt{GCC Fortran}.
If a user needs to use another compiler, the \texttt{Makefile} has to be edited appropriately.
The symbolic link will be created by the \texttt{make install} command.\footnote{By default the link will be created in \texttt{HOME/bin}.}
Once the directory is added to \texttt{PATH}, the code is ready to run.
\begin{minted}[breaklines]{text}
 $ make
 $ make install
 $ echo 'export PATH=$PATH:$HOME/bin' >> $HOME/.bashrc
 $ source $HOME/.bashrc
\end{minted}

\subsection{How to run}
A job submission can be controlled by a Python script, and some sample scripts are prepared in the \texttt{exe} directory.
For example, the NN and 3N matrix elements can be generated with \texttt{NuHamil\_2BME.py} and \texttt{NuHamil\_3BME.py}, respectively.
A Python script generates the corresponding input file for \texttt{NuHamil.exe} and submits the job.
If an user needs to run a job manually, it can be done with
\begin{minted}[breaklines]{text}
 $ NuHamil.exe input.txt
\end{minted}
The ``input.txt'' is the input file based on the Fortran namelist functionality, and the file format is given the following.
\begin{lstlisting}[language=Fortran,basicstyle=\ttfamily]
 &input
    variable1=value
    variable2=value
    ...
 &end
\end{lstlisting}

\subsection{Major parameters}
In the \nuhamil code, there are a number of input parameters.
Here, we list some of the basic input parameters that the users might need to change depending on their requirements.\footnote{Users can contact the author for the parameters not listed here.}
\begin{itemize}
\item{\texttt{rank}: integer, particle number of the system.}
\item{\texttt{hw}: frequency of the HO basis in the unit of MeV. The typical range is $10 \lesssim \texttt{hw} \lesssim 40$.}
\item{\texttt{hw\_target}: target frequency for the frequency conversion technique~\cite{Roth2014}. The parameter is valid if $\texttt{rank} > 2$. To turn off the frequency conversion, set $\texttt{hw\_target}=-1$.}
\item{\texttt{emax}: $e_{\rm max}=\max(2n+l)$ truncation for the output lab-frame HO matrix element file.}
\item{\texttt{e2max}: $e_{\rm 2max}=\max(2n_{p}+l_{p}+2n_{q}+l_{q})$ truncation for the output lab-frame HO matrix element file. It is recommended to use $\texttt{e2max} = 2 \times \texttt{emax}$.}
\item{\texttt{e3max}: $E_{\rm 3max}=\max(2n_{p}+l_{p}+2n_{q}+l_{q}+2n_{r}+l_{r})$ truncation for the output 3N lab-frame HO matrix element file. A typical limit is $\texttt{e3max} = 16$, and it will not work for the larger $\texttt{e3max}$ because of the memory requirements. If only the matrix elements relevant for the NO2B approximation are needed~\cite{Miyagi2022}, $\texttt{e3max}=24$ would be a typical choice without MPI parallelization.}
\item{\texttt{file\_name\_nn}: file name of the output lab-frame NN HO matrix elements.}
\item{\texttt{file\_name\_3n}: file name of the output lab-frame 3N HO matrix elements.}
\item{\texttt{renorm}: renormalization method; ``bare'', ``srg'', and ``Vlowk'' are available. Note that the code does not support the 3N evolution for ``Vlowk'' option.}
\item{\texttt{renorm\_space2}: NN interaction renormalization space; ``ho'' and ``mom'' are available. The NN renormalization procedure is done in HO-(``ho'') or momentum-(``mom'')space. The default is ``ho''.}
\item{\texttt{input\_nn\_file}: file name of the NN interaction represented in the relative momentum space. The files are in the \texttt{input\_nn\_files} directory.}
\item{\texttt{NNInt}: name of the NN interaction.}
\item{\texttt{N2max}: $N_{\rm max}$ truncation for the NN system.}
\item{\texttt{only\_no2b\_element}: If it is set \texttt{True}, only the matrix elements relevant for the NO2B approximation will be computed~\cite{Miyagi2022}.}
\item{\texttt{jmax3}: maximum value of the 3N Jacobi angular momentum taken into account. This has to be an integer and twice the actual angular momentum.}
\item{\texttt{genuine\_3bf}: set \texttt{True} if the bare 3N interaction needs to be included. If it is set \texttt{False} and \texttt{renorm}$=$``srg'', the SRG induced 3N interaction will be computed.}
\item{\texttt{Regulator}: 3N regulator functional form. One can choose ``Local''~\eqref{eq:freg_local},  ``NonLocal''~\eqref{eq:freg_nonlocal}, or   ``LNL''~\cite{Soma2020}.}
\item{\texttt{RegulatorPower}: power of the regulator function, i.e., $n$ in Eqs.~\eqref{eq:freg_nonlocal} and~\eqref{eq:freg_local}.}
\item{\texttt{LECs}: 5 dimensional array providing the low-energy constants appear in N$^{2}$LO 3N interaction in the chiral EFT, $\{c_{1}, c_{3}, c_{4}, c_{D}, c_{E}\}$. Note that $c_{1}$, $c_{3}$, and $c_{4}$ are in units of GeV$^{-1}$, while $c_{D}$ and $c_{E}$ are dimensionless. For more details, see Ref.~\cite{Epelbaum2002}.}
\item{\texttt{lambda\_3nf\_nonlocal}: cutoff of the 3N non-local regulator, $\Lambda_{\rm nonlocal}$ in Eq.~\eqref{eq:freg_nonlocal}, in the unit of MeV.}
\item{\texttt{lambda\_3nf\_local}: cutoff of the 3N local regulator, $\Lambda_{\rm local}$ in Eq.~\eqref{eq:freg_local}, in units of MeV.}
\end{itemize}

\subsection{File format of input NN interactions in relative momentum space}
As mentioned in Sec.~\ref{sec:NN}, some selected NN interaction files are prepared in the \texttt{input\_nn\_files} directory.
Furthermore, one can use their own momentum-space NN interaction.
The file needs to be written with the binary\footnote{For Fortran users, the file should not include any delimiters.}.
The file should begin with listing the following variables:
\begin{minted}[breaklines,breakanywhere]{text}
Number_of_mesh_points
J_max
Number_of_relative_coordinate_channels
\end{minted}
where the 32-bit integers \texttt{Number\_of\_mesh\_points}, \texttt{J\_max}, and \texttt{Number\_of\_relative\_coordinate\_channels} are the size of momentum mesh points, maximum total angular momentum in the relative coordinate, i.e., ${\rm max}(J^{\rm NN}_{\rm rel})$, and the number of $[J^{\rm NN}_{\rm rel}, (-1)^{l}, S, t_{z,p}+t_{z,q}]$ combinations written in the file, respectively.
Then, one needs to write the momentum mesh points and corresponding weights for a quadrature method, which are \texttt{number\_of\_mesh\_points}-dimensional arrays with the 64-bit float:
\begin{minted}[breaklines,breakanywhere]{text}
momentum_mesh_points
weights
\end{minted}
Finally, the momentum-space matrix for each $[J^{\rm NN}_{\rm rel}, (-1)^{l}, S, t_{z,p}+t_{z,q}]$ block should be written in the following way:
\begin{minted}[breaklines,breakanywhere]{text}
Angular_momentum
Parity
Spin
Isospin_z_component
Matrix_dimension
Momentum_space_matrix
... (The same structure is repeated Number_of_relative_coordinate_channels times)
\end{minted}
Here, the 32-bit integers \texttt{Angular\_momentum}, \texttt{Parity}, \texttt{Spin}, \texttt{Isospin\_z\_component}, and \texttt{Matrix\_dimension} correspond to $J^{\rm NN}_{\rm rel}$, $(-1)^{l}$, $S$, $t_{z,p}+t_{z,q}$, and the size of momentum-space matrix, respectively.
Note that $t_{z,p}+t_{z,q}$ can take either of $-1$ (proton-proton), $0$ (proton-neutron), or $1$ (neutron-neutron).
The \texttt{Momentum\_space\_matrix} is the flattened (``Matrix\_dimension'')$^{2}$-dimensional array with the 64-bit float, corresponding to $V^{SJ^{\rm NN}_{\rm rel}}_{l'l} (p',p)$\footnote{For further details, one can see the subroutine \texttt{read\_nn\_mom} in \texttt{NNForce.F90}.}.
Notice that \texttt{Number\_of\_mesh\_points} and \texttt{Matrix\_dimension} are not always the same because \texttt{Matrix\_dimension} is twice of \texttt{Number\_of\_mesh\_points} for spin-triplet coupled channels.

\subsection{File format of non-local 3N matrix elements in Jacobi momentum space}
As mentioned in Sec.~\ref{sec:3N}, external input HDF5 files are needed for the non-local 3N matrix elements.
The HDF5 files need to be prepared for each $[J^{\rm 3N}_{\rm rel}, (-1)^{l_{12}+l_{3}}, T]$ partial waves and placed in the directory \texttt{(directory path)/T3\_$2T$/J3\_$2J^{\rm 3N}_{\rm rel}$/PAR\_$(-1)^{l_{12}+l_{3}}$}, e.g., \path{HOME/3NF_matrix_elements_nonlocal_V/T3_1/J3_1/PAR_1} for the $^{3}$H and $^{3}$He ground-state channel.
Each HDF5 file must include the entries \texttt{Nalpha}, \texttt{Np}, \texttt{Nq}, \texttt{p mesh}, \texttt{q mesh}, \texttt{pw channels}, and \texttt{matrix elements}, written as the dataset type supported in the HDF5 format.
The \texttt{Nalpha} is the 32-bit integer corresponding to the number of $\alpha$ channels.
The objects \texttt{Np} and \texttt{Nq} are also 32-bit integers and the number of momentum mesh points for $p$ and $q$, respectively.
Note that $\alpha$ is defined in Eq.~\eqref{eq:def_alpha}.
The dataset \texttt{p mesh} should at least have \texttt{mesh point} and \texttt{mesh weight} entries, corresponding to the $p$-momentum mesh points and associated weights for a quadrature method, respectively.
The objects \texttt{mesh point} and \texttt{mesh weight} should be \texttt{Np}-dimensional array with the 64-bit float.
The dataset \texttt{q mesh} is the same as \texttt{p mesh} except that it is for $q$.
Regarding the dataset \texttt{pw channels}, it should at least include \texttt{L\_12}, \texttt{S\_12}, \texttt{J\_12}, \texttt{T\_12}, \texttt{l\_3}, and \texttt{2*j\_3} entries, which are $l_{12}$, $s_{12}$, $j_{12}$, $t_{12}$, $l_{3}$, and $2j_{3}$, respectively.
Each entry has to be ``Nalpha''-dimensional array with the 32-bit integer.
Finally, the ``matrix elements'' corresponds to the 3N matrix stored as [\texttt{Nalpha}, \texttt{Nq}, \texttt{Np}, \texttt{Nalpha}, \texttt{Nq}, \texttt{Np}]-dimensional array with the 32-bit float\footnote{Further details can be found in \texttt{NNNFFromFile.F90}}.

\section{Summary and future perspective}\label{sec:Conclusion}
We introduce the \nuhamil code to generate NN and 3N matrix elements.
The jobs can be managed with a simple Python script.
The available NN interactions are LO -- N$^{4}$LO with 500 MeV regulator cutoff by Entem--Machleidt--Nosyk~\cite{Entem2017}, N$^{3}$LO with 500 MeV regulator cutoff by Entem--Machleidt~\cite{Entem2003}, N$^{2}$LO$_{\rm opt}$~\cite{Ekstrom2013}, $N^{2}$LO$_{\rm sat}$~\cite{Ekstrom2015}, and $\Delta$-full EFT series by Gothenburg--Oak Ridge collaboration~\cite{Jiang2020}.
The code can generate locally regulated 3N interactions~\cite{Navratil2007}.
Additional input files are needed for non-locally regulated 3N interactions.
The code also supports the free-space NN and 3N SRG evolution, and the consistent evolution of the other operators is implemented.
The output files can be used for the NCSM calculations with the \texttt{BIGSTICK} code~\cite{Johnson2018} and IMSRG calculations with the \texttt{imsrg++} code~\cite{imsrg++}.

For a comprehensive understanding of nuclear structure, interactions between a nucleus and external field should be addressed.
For example the electromagnetic observables are results of the nucleus-photon interaction and are related with the multipole components of the electromagnetic current operators.
Although we know that higher-order contributions are essential, see Ref.~\cite{Pastore2013,Gysbers2019} for example, the LO current is used in most calculations due to the complexity of the matrix element calculations.
As a future development, we plan to implement the higher-order current operators, including the two-body contributions.

\section*{Acknowledgements}
The \nuhamil code is greatly inspired by the \texttt{manyeff} code by P.~Navr\'atil and \texttt{VRenormalize} in the Computational Environment for Nuclear Structure (CENS) project.
The code uses VODE library by G.~D.~Byrne and S.~Thompson.
The author thanks P.~Navr\'atil, N.~Shimizu, and N.~Tsunoda for the discussions, optimizations, and parallelizations.
The author also thanks P.~Arthuis, A.~Belly, M.~Heinz, B.~S.~Hu, S.~R.~Stroberg, and A.~Tichai for testing the code and useful feedback.
This work was in part supported by JSPS KAKENHI Grant No. JP16J05707, the Program for Leading Graduate Schools, MEXT, Japan, the Deutsche Forschungsgemeinschaft (DFG, German Research Foundation) -- Project-ID 279384907 -- SFB 1245, and the European Research Council (ERC) under the European Union’s Horizon 2020 research and innovation programme (Grant Agreement No.\ 101020842).
Also, the author was affiliated with TRIUMF receiving funding via a contribution through the National Research Council of Canada.
The code was in part developed and tested with an allocation of computing resources on Cedar at WestGrid and Compute Canada, at the J\"ulich Supercomputing Center, and the Oak Cluster at TRIUMF managed by the University of British Columbia department of Advanced Research Research Computing (ARC).

\bibliographystyle{apsrev4-2}

\end{document}